\documentclass{article}
\usepackage{smc}
\usepackage{amsmath}
\usepackage{multirow}
\usepackage{times}
\usepackage{ifpdf}
\usepackage[english]{babel}
\usepackage{cite}
\usepackage{color, xcolor}

\def\papertitle{MULTI-SOURCE CONTRASTIVE LEARNING FROM MUSICAL AUDIO}
\def\firstauthor{Christos Garoufis}
\def\secondauthor{Athanasia Zlatintsi}
\def\thirdauthor{Petros Maragos}

\newif\ifpdf
\ifx\pdfoutput\relax
\else
   \ifcase\pdfoutput
      \pdffalse
   \else
      \pdftrue
\fi

\ifpdf 
  \usepackage[pdftex,
    pdftitle={\papertitle},
    pdfauthor={\firstauthor, \secondauthor, \thirdauthor},
    bookmarksnumbered, 
    pdfstartview=XYZ 
   ]{hyperref}
  \usepackage[pdftex]{graphicx}
  \graphicspath{{./figures/}}
  \DeclareGraphicsExtensions{.pdf,.jpeg,.png}

  \usepackage[figure,table]{hypcap}

\else 
  \usepackage[dvips,
    bookmarksnumbered, 
    pdfstartview=XYZ 
  ]{hyperref}  

  \usepackage[dvips]{epsfig,graphicx}
  \graphicspath{{./figures/}}
  \DeclareGraphicsExtensions{.eps}

  \usepackage[figure,table]{hypcap}
\fi

\hypersetup{
    colorlinks,%
    citecolor=black,%
    filecolor=black,%
    linkcolor=black,%
    urlcolor=black
}

\title{\papertitle}

\oneauthor
   {\firstauthor$^{1,2,3}$, \secondauthor$^{1,2,3}$, \thirdauthor$^{2,3}$ } {
        $^1$Institute of Language and Speech Proc., Athena Research Center, Athens, Greece \\
     $^2$Institute of Robotics, Athena Research Center, Athens, Greece \\
     $^3$School of ECE, National Technical University of Athens, Athens, Greece \\%
     {\tt \small \href{mailto:cgaroufis@mail.ntua.gr}{cgaroufis@mail.ntua.gr}, \href{mailto:nancy.zlatintsi@athenarc.gr}{nancy.zlatintsi@athenarc.gr},\href{mailto:maragos@cs.ntua.gr}{maragos@cs.ntua.gr}}}

\begin{document}
\capstartfalse
\maketitle
\capstarttrue
\begin{abstract}
Contrastive learning constitutes an emerging branch of self- supervised learning that leverages large amounts of unlabeled data, by learning a latent space, where pairs of different views of the same sample are associated. In this paper, we propose musical source association as a pair generation strategy in the context of contrastive music representation learning. To this end, we modify COLA, a widely used contrastive learning audio framework, to learn to associate a song excerpt with a stochastically selected and automatically extracted vocal or instrumental source. We further introduce a novel modification to the contrastive loss to incorporate information about the existence or absence of specific sources. Our experimental evaluation in three different downstream tasks (music auto-tagging, instrument classification and music genre classification) using the publicly available Magna-Tag-A-Tune (MTAT) as a pre-training dataset yields competitive results to existing literature methods, as well as faster network convergence. The results also show that this pre-training method can be steered towards specific features, according to the selected musical source, while also being dependent on the quality of the separated sources.  
\end{abstract}

\section{Introduction}\label{sec:introduction}
Self-supervised learning (SSL)~\cite{dosovitsky15, byol20} has recently emerged as a learning paradigm that can take advantage of large, unlabelled datasets. An approach that has gained traction in multiple domains involves contrastive learning~\cite{vandenoord18, simclr19, moco20, chen21, saeed21, gao21}, where positive pairs of different \textit{views} of samples, created by suitable augmentation pipelines, are projected in a latent space where projections of each pair are enforced to be as close as possible, while deviating from projections of other pairs. This process can be utilized for pre-training neural networks that can then be fine-tuned to a specific task-at-hand, or by using directly the learned representations as feature vectors for supervised downstream tasks.

In the case of music processing, contrastive SSL has proven competitive to fully supervised alternatives in tasks such as music auto-tagging~\cite{spijkervet21, xiao22}, genre recognition~\cite{xiao22} and music recommendation~\cite{choi22} by means of either large music collections~\cite{xiao22}, or sufficiently sized out-of-domain datasets \cite{srivastava22, saeed21}. Moreover, a recent development concerns the transfer of contrastive learning in a multimodal setting, combining information between, i.e., 
audio and text~\cite{manco22, huang22}. 

An important part of every contrastive learning paradigm is the method used to create the positive pairs. Its importance is profound, since it can lead to large variations in the performance of downstream tasks~\cite{simclr19,spijkervet21}. In audio representation learning, the most basic methodology for generating positive pairs is cropping different segments from the same audio clip~\cite{saeed21}, which, in most cases, is also included in more sophisticated augmentation pipelines. In the case of time-frequency representations, inspired by SpecAugment~\cite{specaugment19}, time and frequency masking, pitch shifting, and temporal warping can be applied~\cite{xiao22, emami21, choi22}, whereas augmentation pipelines employed for raw audio may involve polarity shift, artificial reverberation, pitch shifting, gain reduction, or additive noise~\cite{spijkervet21}.

\begin{figure*}[t]
        \centering
        \centerline{\includegraphics[width=17cm]{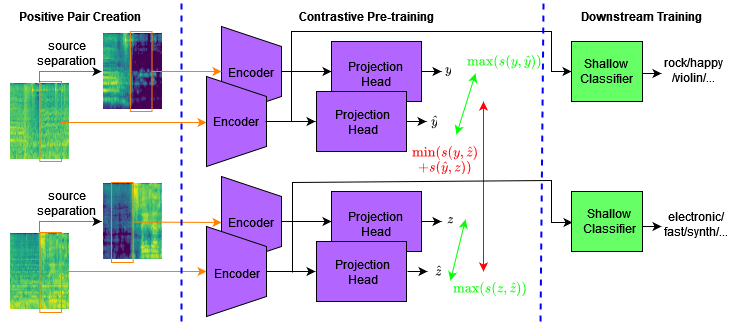}}
    \vspace{-0.4cm}
    \caption{\textcolor{black}{An overview of our proposed framework for learning audio representations, where pairs of song excerpts with a corresponding source are used to pre-train an encoder network through a contrastive loss objective. This enforces the representations of each pair to be close in a latent space, and the representations of different pairs to be further apart.}}
    \label{fig:ovrl}
    \vspace{-0.55cm}
\end{figure*}

\textcolor{black}{Musical pieces usually consist of multiple musical sources, highly coordinated in terms of tonality and rhythm. Besides the harmonic properties of the co-playing sources, the existence or absence of a particular source in a piece also carries semantic information; classical music rarely contains vocals, with the exception of chorals, drums and bass are more prominent in rock or electronic songs, etc.} \textcolor{black}{Finally, each of the co-playing sources conveys distinct information about the genre and rhythmical, or high-level, properties of the musical piece.} Motivated by the above, in this work we propose musical source association (MSA)\footnote{Code/models available:  \href{https://github.com/cgaroufis/MSCOL_SMC23}{https://github.com/cgaroufis/MSCOL\_SMC23}} as a pretext task for audio representation learning, where a neural network learns to associate song excerpts with either a stochastically selected (multi-source) or a particular (source-targeted) vocal or instrumental source. The proposed framework also utilizes semantic information, by taking into account the existence or absence of particular sources in each excerpt. This diverges from a number of research works in the literature that utilize a specific source, such as the singing voice~\cite{lee19,kim21,yakura22} or percussion~\cite{dorian22,heydari22}. Application of our proposed methodology on top of COLA~\cite{saeed21}, a popular self-supervised audio representation learning framework indicates that it can yield competitive results to a number of positive pair creation strategies ~\cite{spijkervet21,xiao22} in three downstream tasks. We further show that models trained to associate song excerpts with a particular source can be steered towards recognizing musical properties tied to this instrumental source. Finally, we repeat the study conducted in~\cite{fonseca21waspaa} concerning the clarity of the separated sources, with the results indicating that in contrast to the case of environmental sounds~\cite{fonseca21waspaa}, the quality of the learned representations correlates with the clarity of the separated musical sources.

The rest of the paper is organized as follows: We discuss the related literature in Sec.~2, and present our proposed methodology in Sec.~3. Sec.~4 describes the datasets used in this work, as well as the preprocessing applied upon them, while in Sec.~5 we outline our experimental setup. The results are presented and discussed, for both multi-source and source-targeted models, in Sec.~6, while we draw some final conclusions in Sec.~7.

\section{Related Work}

During recent years, a multitude of frameworks suited for contrastive SSL from musical signals have appeared in the literature. These frameworks can be decomposed into a number of distinct components: i) the \textit{augmentation pipeline} that is used to generate the positive pair of different views, ii) the \textit{backbone encoder} that learns the intermediate representations, iii) the \textit{projection head}, where the resulting representations are projected, and iv) the \textit{contrastive loss function} that is applied upon the projected representations. 

COLA \cite{saeed21} constitutes a simple, yet effective framework for contrastive audio representation learning in the time-frequency domain, operating on mel-spectrograms. Anchor-positive pairs are generated by cropping different temporal segments of the same mel-spectrogram, and then processed through an EfficientNet-B0~\cite{tan19} backbone, to provide an embedding of fixed length. To formulate a learning objective upon these embeddings, the network is tasked to identify the positives that correspond to each anchor within a batch. Embeddings learned from a sufficiently large pre-training dataset~\cite{gemmeke17}  showed to have high generalization ability across different speech and music tasks.

Numerous works in music representation learning utilize this core methodology, altering some of its components. For instance,~\cite{xiao22} propose utilizing a backbone based on Swin-Transformers~\cite{liu21} in conjunction with a data-driven augmentation pipeline, which involves time cropping, pitch shifting, time warping, and stochastic masking of the spectrograms in both time and frequency axes. Moreover, a number of positive mining strategies and loss function formulations on top of a CNN backbone are compared in~\cite{choi22}. In~\cite{spijkervet21}, the authors successfully applied these principles in the waveform domain, by designing a series of augmentations applied in the waveform level, and then a suitable encoder based on SampleCNN~\cite{lee18} for learning the intermediate representations. Finally, utilization of editorial metadata in the form of information about the artist, album, and others, has also been proposed for pair mining in~\cite{alonso22}, yielding good results in a number of downstream tasks.

Another category of pretext tasks utilized for learning high-level audio representations is based on source separation. In~\cite{fonseca21icassp}, additive mixtures of audio excerpts are used in conjuction with their components as input pairs, which the authors expand on in~\cite{fonseca21waspaa} by using an unsupervised source separation system~\cite{wisdom20} to extract separated views of the initial audio segment. In music signal processing, similarities between song excerpts and the corresponding vocals have been exploited, using triplet losses~\cite{lee19,kim21} or batch-wise contrastive losses~\cite{yakura22}, usually for the task of artist identification. Furthermore, in~\cite{dorian22} the positive pairs are created using the percussive part of a music segment and its non-percussive accompaniment, without temporal cropping -- an idea that was expanded in~\cite{heydari22}. The features learned this way were shown to outperform previous methods in the task of beat synchrony prediction. 

\section{Methodology}

\textcolor{black}{An overview of our proposed approach is displayed in Fig.~1. Building upon the COLA~\cite{saeed21} framework, pairs of song excerpts and a corresponding source are used to pre-train an encoder via a contrastive loss objective. The encoder is tasked to maximize a similarity measure between learned representations of the song excerpt and its separated source, while minimizing it for song excerpts and separated sources from different input songs. Afterwards, these representations can be used for other downstream tasks.}

\subsection{Positive Pair Creation}

We consider the additive mixture model for musical signals, where each signal can be decomposed in $M$ co-playing sources, which can be either active or silent. Thus, we generate positive pairs by matching each song excerpt (further denoted as anchor) with a time-shifted version of an extracted source from the same excerpt (denoted as positive), such as the vocals or one of the instruments. \textcolor{black}{Association of a musical excerpt with an isolated source enables its easier association with sonic or high-level properties tied to a particular source. Furthermore, time-shifting the source excerpt, similar to~\cite{saeed21}, both allows the network to learn long-term dependencies in the audio and acts as data augmentation, increasing the positive pairs seen during training. This framework can be used to} train either models targeted to the association of song excerpts with multiple sources, or a particular, source-targeted one (such as the vocals, the melodic accompaniment, the percussion, etc).

We note that \textcolor{black}{in both cases}, each batch contains \textcolor{black}{source excerpts} from the same musical source; \textcolor{black}{thus, they have similar timbre, increasing the specificity of the learned representations}. Furthermore, in the case of models trained on multiple sources, \textcolor{black}{each batch may either include only non-silent source segments, or source segments sampled from the whole training set, which could contain silence.} Sole utilization of the first approach discards a large portion of training data and prevents the model from learning semantic information \textcolor{black}{pertaining to the existence or absence of a source}. On the other hand, the second approach on its own yields, due to the appearance of multiple silent segments per batch, a smaller effective batch size, which is detrimental to learning useful representations from SSL~\cite{simclr19,saeed21,spijkervet21}.

\subsection{Network Backbone}

Similar to~\cite{saeed21}, the encoder network is based on EfficientNet-B0~\cite{tan19}. In more detail, it consists of a series of inverted residual depthwise-convolutional blocks with linear bottlenecks~\cite{sandler18}, grouped in stages, with each stage gradually reducing the resolution of its input. The output tensor of the last stage is flattened via a global max pooling operation, to acquire an one-dimensional feature vector for each input instance, with fixed size equal to 1280. On top of the encoder, we apply a linear projection head to obtain an embedding vector, of dimensionality equal to 512, followed by a Layer Normalization and a $\mathrm{tanh()}$ activation.

\subsection{Contrastive Loss Objective}

Contrastive learning methods apply the loss function upon the projected representations of their inputs within-batch, i.e., they try to identify the instances that correspond to the respective anchors among other instances in the batch. \textcolor{black}{However, in the cases where the selected source spectrogram is silent, the same positive instance will correspond to multiple anchor spectrograms, generating thus false positives.} Thus, before applying a similarity function between the embedding pairs, we aggregate all embeddings of silent source instances into a single embedding vector, $\bar{y}$. Comparisons are performed within-batch; for pairs where the positive example corresponds to an active source, the embeddings of the anchor are encouraged to be close to the embeddings of their positive pair, while far from those of the other positive instances in each batch. Similarly, for anchor excerpts with a silent source, the embeddings are attracted to the within-batch centroid of the silent positive embeddings, and repelled from all other instances corresponding to active sources. Following~\cite{saeed21}, the bilinear similarity is used as the similarity function between two instances $y, \hat{y}$, defined as:
\vspace{-0.15cm}
\begin{equation}
    s(y,\hat{y}) = y^T W\hat{y},
\end{equation}
\textcolor{black}{where $W$ denotes the corresponding bilinear coefficient matrix}, implemented as an additional learnable linear layer that is applied upon $\hat{y}$. To formulate a loss objective upon the pair similarities, the softmax function is applied upon the batch similarity matrix, with a temperature parameter $T$, and a binary cross-entropy loss function is estimated from the computed logits:
\vspace{-0.15cm}
\begin{equation}
\mathcal{L} = -\sum_{y \in S}{\mathrm{log} \dfrac{\mathrm{exp}( s(y,\hat{y})/T)}{\sum_{z\in S^{*-}} (\mathrm{exp}(s(y,z)/T))}}.
\end{equation}
\textcolor{black}{This loss is similar to the variant of the NT-XEnt loss~\cite{sohn16} used in~\cite{saeed21}, but is modified to accomodate the false positives from including silent source excerpts}. In particular, $S$ denotes the set of anchor embeddings in the current batch, but $S^{*-}$ now denotes the modified distractor embeddings, namely, the intact embeddings of the non-silent positives and the centroid of the embeddings of the silent positives. 

\section{Data and Preprocessing}

\textbf{Datasets}: As our pre-training dataset, we would ideally want to utilize a large dataset of \textcolor{black}{music tracks, paired with their isolated musical sources}. However, no such dataset is publicly avaliable, with the most popular dataset for benchmarking music source separation algorithms~\cite{musdb18} having a total duration of around 10 hours. \textcolor{black}{An intriguing alternative stems from the increase in the performance of automatic audio source separation algorithms in recent years~\cite{luo22, kong21, mdx21, manilow22}, which allows the creation of large datasets consisting of high-quality automatically separated sources. However,} the process of extracting the various sources from large unlabelled music collections is time-consuming. As a result, we utilized the Magna-Tag-A-Tune (MTAT) dataset~\cite{mtat}, which includes 25863 song clips, sampled at 16 kHz, with a duration of 30 seconds each, \textcolor{black}{leading to a total duration of 215 hours. This allows for the acquisition of the estimated source excerpts via automatic source separation algorithms in a sensible time period, while being sufficiently large for self-supervised audio pre-training~\cite{spijkervet21,manco22}}.  

Each of the song clips in MTAT is associated with a number of tags. MTAT contains a total of 188 unique tags, \textcolor{black}{out of which} the top 50 are used as a popular benchmark for music auto-tagging. In this work, we use the filtered version of MTAT presented in~\cite{won20harmonic} which includes \textcolor{black}{only samples associated with any of the 50 most frequent tags.} 

Concerning the downstream tasks, we re-used MTAT for the task of music auto-tagging. \textcolor{black}{Moreover, to examine the cross-dataset generalization of our approach}, we utilized NSynth~\cite{nsynth} for the instrument family classification task, and the \textit{small} subset of the FMA dataset~\cite{fma16} for the task of genre classification. NSynth includes approximately 300000 monophonic note instances from 11 instrument families, sampled at 16 kHz, whereas the \textit{small} subset of the FMA dataset contains 8000 song excerpts, each 30 seconds long, sampled at 16 kHz, and equally split into 8 root genres.  

\textbf{Data Preprocessing}: All audio segments corresponding to MTAT were sliced into 4-sec chunks, and mel-spec-trograms were extracted using a window size of 25ms, a hop length of 10ms, and 64 mel bands, following~\cite{saeed21}. Since in~\cite{saeed21} an input segment length of 1 second is utilized, random segments of 98 frames were cropped, at each training iteration, for both anchor and source samples.

 To acquire the audio excerpts corresponding to various musical sources, we used the open-source open-unmix~\cite{stoter19} framework for music source separation, which decomposes a musical mixture into the vocal track, the bass track, the drums, and the melodic accompaniment. Open-unmix operates on the time-frequency domain, and specifically in the STFT magnitude, estimating soft masks for each source, which are then applied in the STFT of the mixture to isolate them. The song excerpts were upsampled to 44.1~kHz for compatibility purposes, and the extracted sources were downsampled to 16~kHz before spectrogram computation. As a further pre-processing step, \textcolor{black}{after inspecting the energy of the extracted sources}, we replaced all source excerpts with a mean absolute value below 0.01 with silence. We observe that with the exception of the melodic accompaniment (``Other''), the rest of the sources are active for less than half of the audio segments; \textcolor{black}{in particular, 44.67 \% of the bass segments, 41.94 \% of the drum segments, 87.72 \% of the melodic accompaniment segments and 31.50 \% of the vocal segments in the training set are active.}

 \section{Experimental Protocol}

The training procedure was split into two stages: a) self-supervised pre-training of the encoder along with the projection head and b) training of a shallow classifier upon the learned encoded representations (freezing the encoder and dropping the projection head). Models were trained for both source-targeted (single-source) and multi-source cases. In the former case, the positive pairs were created only from one extracted source, while in the latter case, a source was selected at random for each batch. In both cases, the encoder and the projection head were trained via the contrastive loss objective defined in Eq.~(2), using $T=0.2$. The training procedure lasted 10000 steps (approximately 800 epochs for MTAT), with each step consisting of 64 batches; due to GPU limitations, each batch consisted of 128 positive pairs. Adam was used for the self-supervised pre-training with an initial learning rate equal to 0.001, which was halved after 5000 steps. During pre-training, we only use the training subset of MTAT\footnote{According to the split in  \href{https://github.com/jongpillee/music_dataset_split}{https://github.com/jongpillee/music\_dataset\_split}}, corresponding to unseen testing data. The pretext task performance at the validation set was used for early stopping; we halted pre-training after the running average of the validation loss over 1000 steps did not improve.

\begin{table*}[t]
    \begin{center}
    \begin{tabular}{|c||c|c||c|c||c||c|} \hline

   SSL Framework & \multicolumn{2}{c||}{MTAT} & \multicolumn{2}{c||}{MTAT$^{*}$} & NSynth & FMA \\ \cline{2-7}  & ROC-AUC & PR-AUC & ROC-AUC & PR-AUC & WA (\%) & WA (\%)\\ \hline
    \hline
        CLMR~\cite{spijkervet21} & - & - & 0.887 & 0.356 & - & 0.484 \\
    \hline \hline
          COLA~\cite{saeed21} & 0.886 & 0.396 & 0.880  & 0.334 & 0.593 & 0.460\\
     \hline
           COLA~\cite{saeed21} + MWS~\cite{xiao22} & 0.898 & 0.425 & 0.892 & 0.358 & \textbf{0.645} & 0.493 \\ 
     \hline
        COLA~\cite{saeed21} + Random Mask & 0.883 & 0.390 & 0.880 & 0.337 & 0.632 & 0.476 \\ \hline
        COLA~\cite{saeed21} + MSA (ours) & \textbf{0.900} & \textbf{0.429} & \textbf{0.895} & \textbf{0.361} & 0.627 & \textbf{0.510} \\
     \hline
    \end{tabular}
    \vspace{-0.1cm}
            \caption{\textcolor{black}{Performance of the pre-trained representations} in MTAT in different downstream tasks: music auto-tagging (MTAT and MTAT$^*$), instrument classification (NSynth) and genre classification (FMA), according to the strategy used for generating the positive pairs. The results for CLMR are collected from~\cite{spijkervet21} and ~\cite{xiao22}.}
    \end{center}
    \vspace{-0.15cm}
    \label{tab:compare}
    \vspace{-0.5cm}
\end{table*} 

Concerning the training of the downstream classifiers, in the case of MTAT, the common 12:1:3 split between training, validation and testing data~\cite{won20protocol,spijkervet21} was employed, \textcolor{black}{while for both NSynth and FMA we utilized the default splits between training, validation and testing data: in the case of NSynth, the training, validation and testing sets consist of 289205, 12678, and 4096 audio segments, while for FMA, the data are split in a stratified 8:1:1 ratio into the respective splits.} For the cases of MTAT and FMA, we directly used a linear classifier, while for NSynth, we used an intermediate layer with 512 neurons and a ReLU activation function, as in~\cite{manco22,xavier20}. The classifiers were trained using Adam with a learning rate of 0.0005 for MTAT and FMA and 0.0003 for NSynth, and a batch size of 128, while early stopping was applied with a patience of 5 epochs. 
Since music auto-tagging constitutes a multi-instance multi-label task, we use the categorical cross-entropy as its loss function, and report on the macro average ROC-AUC and PR-AUC values over all tags. On the other hand, for both instrument family classification and music genre classification, the binary cross-entropy is used as the loss function and the weighted accuracy (WA, \%) as the evaluation metric. During evaluation, all models receive 1-sec audio segments, and aggregate the predictions for each instance via averaging the per-segment predictions. \textcolor{black}{We further note that all downstream classification experiments were repeated 5 times, and we report on their average scores. To assess the statistical significance of the results, t-tests were performed between the results of the repeated experiments.}

\begin{table*}
    \begin{center}
    \begin{tabular}{|c||c|c||c|c||c||c|} \hline

   Configuration & \multicolumn{2}{c||}{MTAT} & \multicolumn{2}{c||}{MTAT$^{*}$} & NSynth & FMA \\ \cline{2-7}  & ROC-AUC & PR-AUC & ROC-AUC & PR-AUC & WA (\%) & WA (\%)\\ \hline
    \hline
        COLA \cite{saeed21} + MSA  & \textbf{0.900} & \textbf{0.429} & \textbf{0.895} & \textbf{0.361} & 0.627 & \textbf{0.510}  \\
    \hline \hline
              NSV1 & 0.896 & 0.418 & 0.891  & 0.351 & 0.648 & 0.501 \\
     \hline
               NSV2 & 0.895 & 0.418 & 0.890  & 0.354 & 0.645 & 0.486 \\
     \hline \hline
          A1 & 0.898 & 0.424 & 0.893  & 0.358 & 0.635 & 0.499 \\
     \hline
           A2 & 0.897 & 0.423 & 0.892 & 0.357 & \textbf{0.653} & 0.504\\ 
     \hline
        A3 & 0.841 & 0.310 & 0.838 & 0.262 & 0.537 & 0.425 \\
     \hline
    \end{tabular}
    \vspace{-0.2cm}
            \caption{Ablation study on the effect of the batch composition on the performance in different downstream tasks.}
    \end{center}
    \label{tab:ablate}
    \vspace{-0.9cm}
\end{table*} 

We compared our proposed methodology (MSA, music source association) against the following baselines, which we retrain using the same pre-training dataset:
\vspace{-0.2cm}
\begin{itemize}
  \setlength\itemsep{0.4em}
    \item The unmodified COLA~\cite{saeed21} framework, where the positive pairs were created by choosing different segments of the same audio clip.
    \item The encoder, projection and contrastive loss used in~\cite{saeed21}, coupled with the data-driven methodology to create positive pairs described in~\cite{xiao22}, denoted here as MWS (masking, warping and shifting).
    \item The proposed methodology, \textcolor{black}{but instead of using a fully trained source extractor for obtaining the positive instances}, the various sources are extracted from a randomly initialized, untrained variant of open-unmix~\cite{stoter19}. This way, we investigate whether targeted masking of the input helps in learning useful representations as opposed to random soft masking.
  
\end{itemize}
\vspace{-0.15cm}  
Finally, we compare our method to CLMR~\cite{spijkervet21}, a wave-form-based contrastive learning framework for music, which was also trained on MTAT as a pre-training dataset. 

\section{Results and Discussion}

\subsection{Multi-Source Models}

\textbf{Comparison to Baselines}: The performance of our proposed method in all three downstream tasks, in comparison to the baselines mentioned above, is presented in Table~1. For easier comparison with the literature, we report results on both filtered (MTAT) and unfiltered (MTAT$^*$) versions of MTAT. We observe that MSA outperforms both the COLA baseline~\cite{saeed21} and application of random soft binary masks (COLA + Random Mask) on both versions of MTAT, at a statistical significance level of $p < 0.01$ for all metrics. \textcolor{black}{In fact, the performance of the random soft masking variant of MSA is similar to COLA ($p > 0.005$ for all metrics after Bonferroni correction), indicating that unstructured masking of the spectrograms on its own cannot provide sufficient information towards learning useful intermediate representations.} Furthermore, we note that our method performs comparably ($p >$ 0.005 for all metrics post-Bonferroni) to the data-driven MWS methodology for creating the positive pairs, even though we employ a simpler augmentation pipeline -- merely matching an audio excerpt to a cropped segment of an extracted source. In the cross-dataset setup, MSA only outperforms COLA ($p<0.01$) in the task of instrument classification, and both COLA and COLA+Random Mask ($p<0.01$) in genre recognition, with no statistically significant difference with MWS ($p > 0.05$). In comparison to CLMR~\cite{spijkervet21}, we observe that we obtain better results on FMA, while achieving comparable performance on MTAT$^*$. Finally, we note that the performance gap between MTAT and MTAT$^*$ is consistent with the literature~\cite{alonso22}.

\textbf{Ablation Study}: In order to investigate the properties of our proposed framework, \textcolor{black}{we perform an extensive ablation study. The effect of incorporating silent source excerpts during training is examined through the following variants:}
\begin{itemize}
\vspace{-0.2cm}
  \setlength\itemsep{0.2em}
    \item Non-Silent Variant \#1 (NSV1): All batches contain non-silent positive examples from the same source, and silent source excerpts are discarded.
    \item Non-Silent Variant \#2 (NSV2): Similar to NSV1, but the source segments in each batch may correspond to different sources.
\end{itemize}
\vspace{-0.15cm}
\textcolor{black}{Furthermore, the importance of performing time-shifting to the source excerpts, modifying the loss function to incorporate silent segments, and constructing training batches containing a single source per batch, is examined through the following ablated variants, using the full training dataset:}
\begin{itemize}
\vspace{-0.2cm}
  \setlength\itemsep{0.2em}
    \item Ablation \#1 (A1): Batches consist of positive pairs that contain multiple, different sources in each batch.
    \item Ablation \#2 (A2): The loss function of ~\cite{saeed21} is used without the modification applied to accomodate the false positives in the case of silent segments.
    \item Ablation \#3 (A3): The source excerpts are temporally aligned with the corresponding song excerpt.
\end{itemize}
    \vspace{-0.15cm}

\begin{table*}
    \begin{center}
    \begin{tabular}{|c||c|c||c|c||c||c|} \hline

   Source & \multicolumn{2}{c||}{MTAT} & \multicolumn{2}{c||}{MTAT$^{*}$} & NSynth & FMA \\ \cline{2-7}  & ROC-AUC & PR-AUC & ROC-AUC & PR-AUC & WA (\%) & WA (\%)\\ \hline
    \hline
        Bass & 0.885 & 0.392 & 0.880 & 0.332 & 0.624 & 0.487 \\
    \hline
        Drums & 0.881 & 0.385 & 0.876 & 0.328 & 0.611 & \textbf{0.503} \\
     \hline
           Accomp. & 0.892 & 0.410 & 0.888 & \textbf{0.348} & \textbf{0.636} & 0.480 \\ 
     \hline
        Vocals & \textbf{0.896} & \textbf{0.412} & \textbf{0.891} & \textbf{0.348} & 0.632 & 0.481 \\
     \hline \hline
             None (COLA) & 0.886 & 0.396 & 0.880 & 0.334 & 0.593 & 0.460\\ \hline
             Multi-Source & 0.900 & 0.429 & 0.895 & 0.361 & 0.627 & 0.510\\
     \hline
    \end{tabular}
    \vspace{-0.15cm}
            \caption{\textcolor{black}{Performance of the pre-trained representations (from source-targeted models)} in MTAT on different downstream tasks, according to the source used for generating the positive pairs.}
    \end{center}
    \label{tab:unisource}
    \vspace{-0.9cm}
\end{table*} 

The results of this ablation study are displayed in Table~2. We note that the proposed framework significantly outperforms all non-silent variants at the $p < 0.001$ level (after Bonferroni correction) in music auto-tagging, and NSV2 ($p < 0.01$ post-Bonferroni) in genre recognition, indicating that incorporating silent segments assists in learning useful features for these tasks. Compared to A1 and A2, MSA achieves slightly higher (but not significant at the $p<0.05$ level) performance in these tasks, implying a slight positive effect of the examined parameters. However, this trend is reversed in instrument classification, with all examined variants, with the exception of A3, scoring at least comparably to MSA, and A2 outperforming it at the $p < 0.05$ level, indicating that the proposed loss is not suitable for this task. Finally, A3 performs significantly ($p<0.001$) worse than MSA in all downstream tasks, showing that time-shifting of the source excerpts plays a critical role in learning informative representations.
\begin{figure}[t!]
\begin{minipage}{0.49\linewidth}
        \centering
        \centerline{\includegraphics[width=4.1cm]{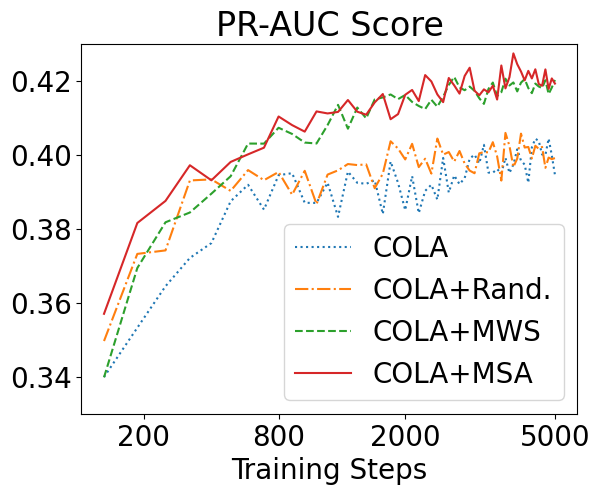}}
    \end{minipage}
    \begin{minipage}{0.49\linewidth}
        \centering
        \centerline{\includegraphics[width=4.1cm]{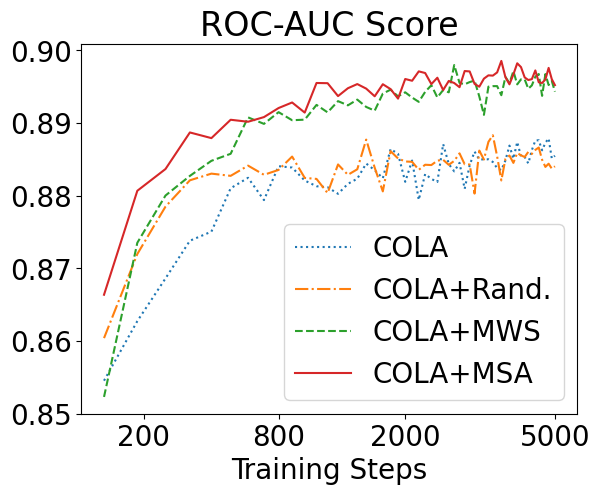}}
\end{minipage}
    \vspace{-0.3cm}
    \caption{\textcolor{black}{Evolution of the PR-AUC and ROC-AUC scores on MTAT in the course of the self-supervised pre-training.}}
    \vspace{-0.35cm}
    \label{fig:evol}
\end{figure}
 \begin{figure}
\begin{minipage}{0.49\linewidth}
        \centering
        \centerline{\includegraphics[width=4.1cm]{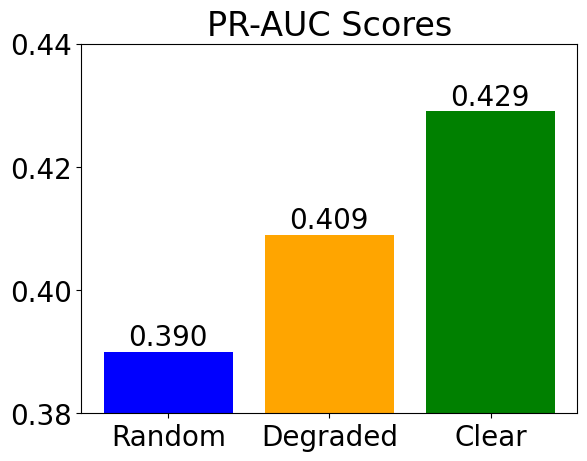}}
    \end{minipage}
    \begin{minipage}{0.49\linewidth}
        \centering
\centerline{\includegraphics[width=4.1cm]{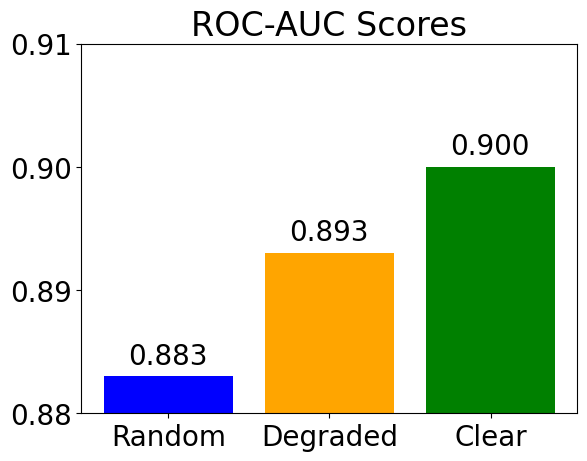}}
\end{minipage}
    \vspace{-0.3cm}
    \caption{\textcolor{black}{PR-AUC (left) and ROC-AUC (right) scores in MTAT, for positive pairs using random (blue), degraded (orange) and clear (green) source excerpts.}}
    \vspace{-0.6cm}
    \label{fig:sourceq}
\end{figure}

\textbf{Qualitiative Analysis}: The evolution of the ROC-AUC and PR-AUC scores in the downstream task of music auto-tagging in MTAT, as evaluated using embeddings from various checkpoints of the pre-training process, is shown in Fig.~\ref{fig:evol}. Interestingly, we note that our proposed methodology (red line) increases its performance faster than its data-driven counterpart~\cite{xiao22} (green dashed line) at the early stages of pre-training, reaching a PR-AUC score of 0.40 within 300 training steps. We assume that the musically motivated pretext task of music source association allows the network to quickly learn meaningful intermediate representations. However, the multitude of generated pairs of the data-driven MWS strategy, due to the inherent stochasticity of the masking, pitch shifting and time warping operations (whereas only 4 different spectrogram excerpts, one for each source, can be cropped in MSA), allows its performance to catch up as pre-training progresses. \textcolor{black}{This could imply that the two strategies can be combined, by utilizing them either in tandem or in succession.}

To further investigate the necessity of utilizing separated sources of high quality, we conducted an experiment similar to~\cite{fonseca21waspaa} to examine the performance of MSA under degrading sources. Thus, for each source, we trained an instance of open-unmix~\cite{stoter19} for just 2 epochs, using the musdb18~\cite{musdb18} dataset.  The ROC-AUC and PR-AUC metrics in MTAT when using the noisy source excerpts, compared to both the completely random masks (COLA + Random Mask) and the clear sources extracted from the fully trained one, are displayed in Fig.~\ref{fig:sourceq}. We observe that in contrast to the case of unsupervised source separation of environmental sounds~\cite{fonseca21waspaa}, the pretext task of source association in musical signals is aided by high quality source excerpts; even though noisy sources lead to worse results ($p < 0.001$) than the output of the fully-trained open-unmix, they still outperform ($p < 0.01$) random masking.
\begin{figure*}[t]
    \vspace{-0.1cm}
        \centering
        \centerline{\includegraphics[width=0.85\linewidth]{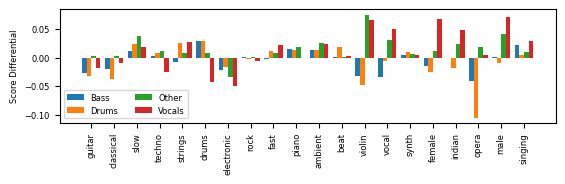}}
    \vspace{-0.7cm}
    \caption{\textcolor{black}{Relative difference (PR-AUC) between the performance of embeddings derived from source-targeted models and COLA-derived embeddings, in the 20 most frequent tags from the MTAT dataset.}}
    \label{fig:aaa}
    \vspace{-0.55cm}
\end{figure*}

\subsection{Source-Targeted Models}

 The results of the models trained to learn the association between a song excerpt and a particular source are displayed in Table~3. We observe that, on average, the models trained using the vocals and the melodic accompaniment perform the best, with a PR-AUC score of 0.412 and 0.410 on MTAT, respectively, outperforming the COLA baseline at the $p<0.05$ level. On the other hand, the bass- and drums-based models do not reach this performance, recording scores slightly lower than the COLA baseline in MTT and slightly higher in instrument classification, with the exception of the genre recognition task, where the drums-based model outperforms (albeit not on the $p < 0.05$ statistical significance level) all other source-targeted models. However, with the exception of the instrument classification task, none of these targeted models reach the performance of the multi-source models, indicating that utilization of multiple sources can be useful for the identification of various low or high-level musical traits.

Indeed, all sources contribute to the performance of the multi-source model, since particular spectral, timbral or higher-level characteristics are primarily associated with each source. To visualize this, in Fig.~\ref{fig:aaa}, we display, \textcolor{black}{for the 20 most frequently appearing tags in} MTAT, the differential in the per-tag PR-AUC scores between COLA~\cite{saeed21} and the models trained with each of all 4 extracted sources: bass, drums, the rest of the melodic accompaniment and vocals. We observe that source-targeted models display a specialization, according to the utilized source. For instance, despite their overall low performance, the embeddings trained to associate song excerpts with the corresponding percussion achieve relatively good performance in rhythmic tags, such as \textit{fast} and \textit{drums}, as well as related genres such as \textit{techno}. In the same vein, embeddings trained on the melodic accompaniment display the highest tagging metrics in the recognition of accompanying instruments, such as \textit{guitar}, \textit{piano}, and \textit{violin}, \textcolor{black}{while those trained using the corresponding vocals perform the best in the related tags \textit{vocal}, \textit{male}, \textit{female} and \textit{singing}}.
Interestingly, none of the source-targeted models outperform the COLA baseline in a number of tags, such as \textcolor{black}{\textit{electronic}}.

\section{Conclusions}

In this paper, we investigated the suitability of musical source association as a pretext task in the context of contrastive music representation learning. Results on three different downstream tasks indicate that our framework can yield competitive results compared to recent literature methods for creating positive pairs under the same pre-training dataset. Furthermore, we show that it can be used to steer downstream classifiers toward specific sonic properties, by using a single (vocal or instrumental) source for the association task. In the future, we would like to explore the scalability of our findings in larger music collections, as well as experiment with other musically motivated pretext tasks, such as mining positive pairs from playlists. Another interesting direction concerns examining the efficiency of our approach in producing genre or high-level tags, using single-track audio, i.e. the vocal track.

\begin{acknowledgments}
This research was supported by the Hellenic Foundation for Research and Innovation (H.F.R.I.) under the “3rd Call for H.F.R.I. Research Projects to support Post-Doctoral Researchers” (Project Number: 7773). \textcolor{black}{We would also like to thank the anonymous reviewers, whose constructive comments helped in improving the quality of this paper.}
\end{acknowledgments}

\bibliography{smc2023template}

\end{document}